\documentclass[aps,prl,reprint,preprintnumbers,superscriptaddress,amsmath,amssymb,bibnotes,longbibliography]{revtex4-2}

\usepackage{graphicx}
\usepackage{dcolumn}
\usepackage{bm}
\usepackage[colorlinks,linkcolor=blue,anchorcolor=blue,citecolor=blue,urlcolor=blue,filecolor=blue,menucolor=blue,runcolor=blue]{hyperref}
\usepackage{ulem}

\begin{document}
\title{Andreev Reflection to Probe Momentum-Dependent Spin Polarization in Altermagnet CrSb}

\author{Yan Zhang}
 \affiliation{Center for Correlated Matter and School of Physics, Zhejiang University, Hangzhou 310058, China}%
 \author{Yixuan Luo}
 \affiliation{State Key Laboratory of Quantum Functional Materials, School of Physical Science and Technology, ShanghaiTech University, Shanghai 201210, China}%
\author{Yue Yang}
 \affiliation{Center for Correlated Matter and School of Physics, Zhejiang University, Hangzhou 310058, China}%
\author{Zilong Li}
\affiliation{Center for Correlated Matter and School of Physics, Zhejiang University, Hangzhou 310058, China}%
\author{Weilong Qiu}
\author{Lunhui Hu}%
\affiliation{Center for Correlated Matter and School of Physics, Zhejiang University, Hangzhou 310058, China}%
\author{Yuanfeng Xu}%
 \email{y.xu@zju.edu.cn}
\affiliation{Center for Correlated Matter and School of Physics, Zhejiang University, Hangzhou 310058, China}%
\author{Yanfeng Guo}   
\email{guoyf@shanghaitech.edu.cn}
\affiliation{State Key Laboratory of Quantum Functional Materials, School of Physical Science and Technology, ShanghaiTech University, Shanghai 201210, China}
\affiliation{ShanghaiTech Laboratory for Topological Physics, ShanghaiTech University, Shanghai 201210, China}
\author{Chao Cao}%
 \email{ccao@zju.edu.cn}
\affiliation{Center for Correlated Matter and School of Physics, Zhejiang University, Hangzhou 310058, China}%
\author{Xin Lu}%
 \email{xinluphy@zju.edu.cn}
\affiliation{Center for Correlated Matter and School of Physics, Zhejiang University, Hangzhou 310058, China}%
\affiliation{Institute of Fundamental and Transdisciplinary Research, Zhejiang University, Hangzhou 310058, China}

\date{\today}

\begin{abstract}

Altermagnetic materials have recently emerged as promising candidates for next-generation spintronic applications, characterized by the $\mathbf{k}$-dependent spin-splitted band structure and a simultaneous zero-net-magnetization. Among them, altermagnetic candidate CrSb has attracted considerable attention, owing to its $g$-wave spin splitting and high N\'eel temperature. In this article, we employed mechanical point-contact spectroscopy (MPCS) with superconducting Nb tips to probe the Andreev reflection on CrSb single crystals along three principal crystallographic orientations. The extracted momentum-dependent spin polarizations are approximately 73.4\% for the (0\,0\,0\,1) plane, 67.9\% for the ($\overline{1}$\,$\overline{1}$\,2\,0) plane, and 61.9\% for the (1\,0\,$\overline{1}$\,0) plane, respectively, distinct from conventional antiferromagnets. Furthermore, conductance spectra from spatial line-scans on the sample surface support the existence of altermagnetic domains with a characteristic size of 250-500~nm separated by domain-walls with width about 250~nm. These results strongly support the momentum-dependent spin polarization in altermagnetic CrSb and establish Andreev reflection as a new paradigm to probe $\mathbf{k}$-dependent spin textures.

\end{abstract}

\maketitle


Spintronics has emerged into a major branch of modern condensed-matter research, which exploits the spin degree of freedom to improve data storage and computing performance  \cite{zuticspintronicsfundamentalsapplications2004,baderspintronics2010a,hirohatareviewspintronicsprinciples2020a,baltzantiferromagneticspintronics2018,lindersuperconductingspintronics2015a,chandrasekarspintronicsminireview2019,fertnobellectureorigin2008,dienyopportunitieschallengesspintronics2020}. The spin polarization $P$ is a crucial parameter to describe the global spin-imbalance between the spin-up and -down bands in spintronics, and can be generally defined as $P = \frac{N_{\uparrow}(E_F) - N_{\downarrow}(E_F)}{N_{\uparrow}(E_F) + N_{\downarrow}(E_F)}$, where $N_{\sigma}(E_F)$ denotes the spin-dependent density of states at the Fermi level \cite{MeasuringPCS}. A finite spin polarization $P$ in ferromagents enables their application in spintronics, while a zero-net-magnetization in antiferromagnets restricts their usage. With the continuing exploration of functional materials, a new class of altermagnetic materials has recently been proposed as an ideal candidate for spintronics \cite{smejkalconventionalferromagnetismantiferromagnetism2022c, smejkalemergingresearchlandscape2022a, rialaltermagneticvariantsthin2024, reichlovaobservationspontaneousanomalous2024a, gottschilchstudyantiferromagnetismmn5si32012,PhysRevB.75.115103, fedchenkoobservationtimereversalsymmetry2024, zhoucrystalthermaltransport2024, tschirnerSaturationAnomalousHall2023a, smejkalchiralmagnonsaltermagnetic2023, smejkalemergingresearchlandscape2022a, leebrokenkramersdegeneracy2024, osumiobservationgiantband2024, guospinsplitcollinearantiferromagnets2023a}. Although the overall spin-polarization $P$ vanishes in altermagnets, the system exhibits a momentum-dependent spin splitting and a new parameter should be introduced to characterize the degree of such momentum-resolved spin polarization as $P_k = \frac{\int d^3k\, |N_{\mathbf{k}, \uparrow} - N_{\mathbf{k}, \downarrow}|}{\int d^3k\, (N_{\mathbf{k}, \uparrow} + N_{\mathbf{k}, \downarrow})}$, where $N_{\mathbf{k}, \sigma}$ denotes the spin- and momentum-resolved density of states at Fermi level. It is thus interesting to experimentally probe the $\mathbf{k}$-dependent spin polarization $P_k$ in altermagnets to distinguish them from conventional ferromagnets or antiferromagnets.

\begin{figure}
  \centering
  \includegraphics[angle=0,width=0.45 \textwidth]{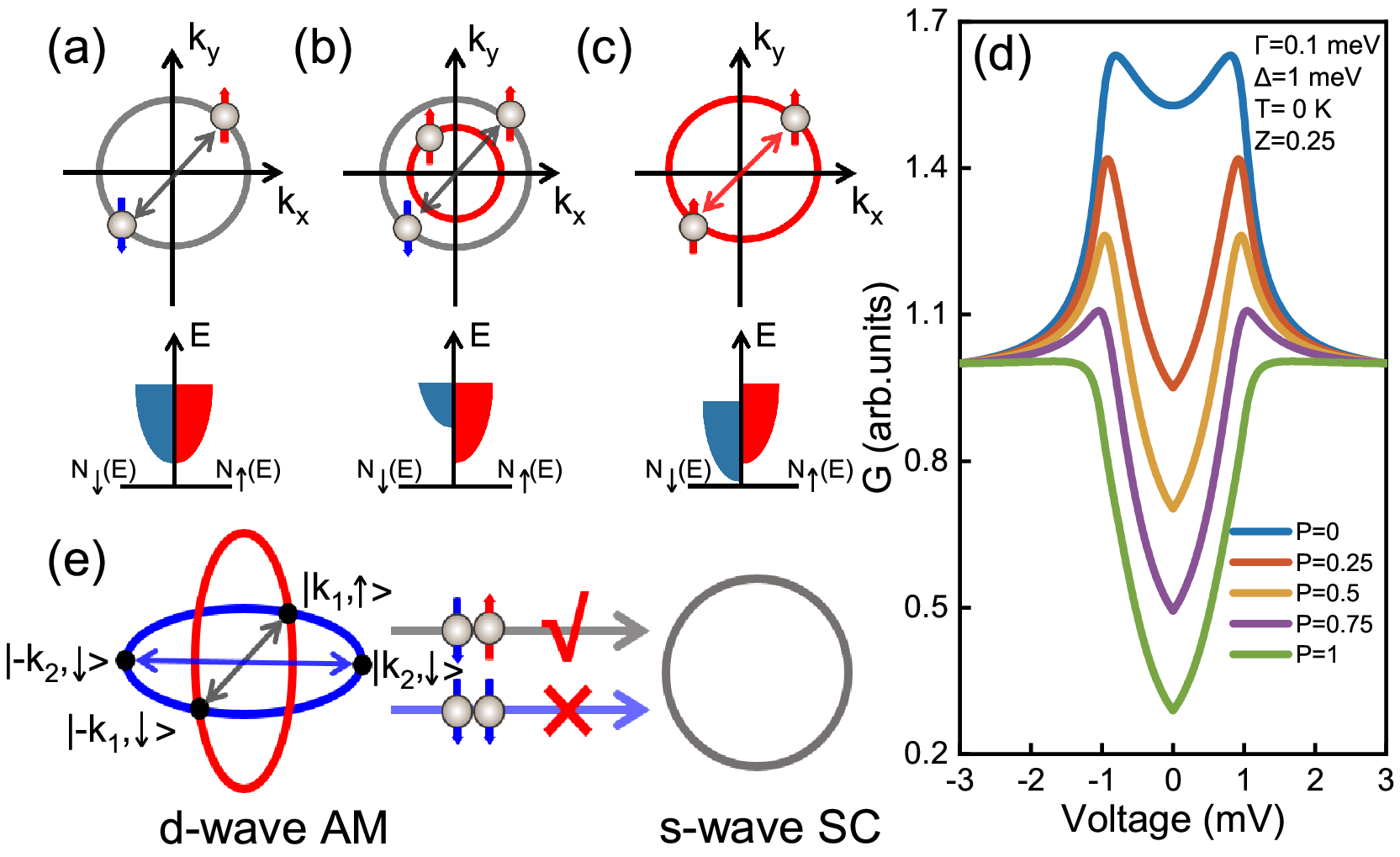}
	\vspace{1mm}
  \caption{
    \textbf{(a)} Normal metal with $P=0$ has a spin-degenerate Fermi surface and symmetric density of states.
    \textbf{(b)} Conventional ferromagnet with $0<P<1$ has one spin-polarized and one spin-degenerate band, leading to an asymmetric density of states at $E_F$.  
    \textbf{(c)} Half-metallic ferromagnet with $P=1$ has only one spin-up band crossing the Fermi level.  
    \textbf{(d)} Simulated conductance curves for MPCS with different spin polarizations.
    \textbf{(e)} Schematic illustration of Andreev reflection at the interface between a d--wave altermagnet and s--wave superconductor. The electrons at $\mathbf{k}$ and $\mathbf{-k}$ have the same spin except for the nodal region, resulting in the suppression of Andreev reflection across the interface. 
 }
  \label{fig1}
	\vspace{-10pt}
\end{figure}

\begin{figure*}
  \includegraphics[angle=0,width=1\textwidth]{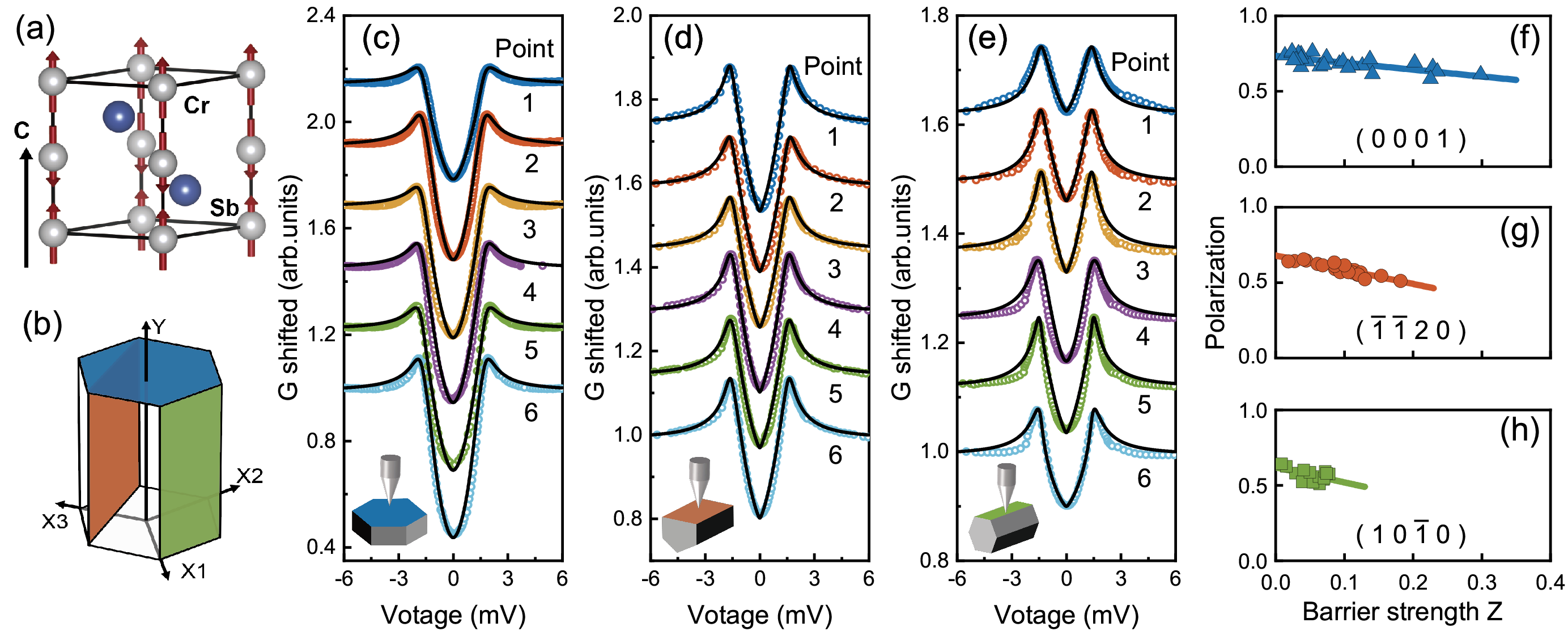}
  \caption{
  \textbf{(a)} Schematic illustration of the crystal and magnetic structure of CrSb. 
  \textbf{(b)} Different crystallographic planes for CrSb: blue (0 0 0 1), orange ($\overline{1}$ $\overline{1}$  2 0), and green (1 0 $\overline{1}$ 0).
  \textbf{(c)} Differential conductance curves $G(V)$ for six point contacts on CrSb crystals along the (0 0 0 1) direction, together with their optimal fits. 
  \textbf{(d)} ($\overline{1}$ $\overline{1}$ 2 0) plane, \textbf{(e)}(1 0 $\overline{1}$ 0) plane. The insets represent the MPCS configuration, where point-contacts were formed between the Nb tip and sample surface for each crystallographic plane. 
  \textbf{(f)-(h)} momentum-dependent spin polarization $P_k$ versus barrier strength Z for different contacts along three major orientations. The dashed lines are a guide to the eye.
  }
    \label{fig2}
		\vspace{-10pt}
\end{figure*}

A prototypical altermagnetic candidate CrSb crystallizes in the hexagonal structure with space group $P6_3/mmc$ as shown in Fig.~\ref{fig2}(a) and its spin symmetry is proposed to be $g$-wave, exhibiting a sixfold spin texture in the $\mathbf{k}$-space \cite{dinglargebandsplitting2024}. In its magnetic ordered state, the Cr moments align ferromagnetically within each plane and antiferromagnetically between adjacent layers separated by nonmagnetic Sb atoms. Neutron diffraction has confirmed its high N\'eel temperature ($T_N = 723$~K) \cite{snowneutrondiffractioninvestigation1952b}, while ARPES measurements on both thin films and bulk crystals have directly observed the large spin-induced band splitting \cite{reimersdirectobservationaltermagnetic2024, yangthreedimensionalmappingaltermagnetic2025a, zengobservationspinsplitting2024, dinglargebandsplitting2024, momentumdependentCrSbArpes2025}. However, the reported spin polarization value in ARPES is significantly smaller and this discrepancy can be ascribed to a large spot size in ARPES. If the light spot covers a large surface area for CrSb with multiple altermagnetic domains in opposite spins, the finite spin-polarization $P$ averages out owing to opposite spins with a domain-averaged effect, similar to the report in MnTe~\cite{krempasky2024altermagnetic}. It is thus desirable to develop an alternative method to access the $\mathbf{k}$-dependent spin polarization with reduced domain-averaging effects.

Point-contact spectroscopy (PCS) and associated Andreev reflection have already been recognized as a reliable method to measure the spin polarization in ferromagnets \cite{MeasuringPCS,upadhyayprobingferromagnetsandreev1998, strijkersAndreevReflectionsMetal2001c, borisovhighfermilevelspin2016, singhcemnni2006}. In the Andreev reflection process, a spin-up (down) electron from the normal metal is retro-reflected as a spin-down (up) hole and a Cooper pair can go across the interface and enter into superconductor, leading to a doubly-enhanced subgap conductance relative to the normal-state value \cite{blondertransitionmetallictunneling1982b}. In the case of ferromagnets, the imbalance between spin-up and -down bands would result in a suppression of Andreev reflection: As shown in Fig.~~\ref{fig1}(a)-(c), when the band spin polarization $P$ increases from 0 to 1, respectively, Andreev reflection is progressively suppressed, yielding a deepened zero-bias dip and diminished coherence peaks in $G(V)$, as in Fig.~\ref{fig1}(d). Its $P$ can be extracted by fitting the conductance spectra $G(V)$ with a modified Blonder--Tinkham--Klapwijk (BTK) model by $G(V) = (1 - P) G_{\text{u}}(V) + P*G_{\text{p}}(V)$ \cite{MeasuringPCS}, where $G_{\text{u}}$ and $G_{\text{p}}$ denotes contributions from spin-unpolarized and -polarized parts, respectively. For ferromagnets, the isotropic spin splitting in $\mathbf{k}$-space guarantees the global spin polarization $P$ identical to $P_{\mathbf{k}}$. However, this equivalence no longer necessarily holds for the case of altermagnets. For example, a d-wave altermagnet exhibits a finite $\mathbf{k}$-dependent spin polarization $P_{\mathbf{k}}$ but a zero global $P$ as in Fig.~\ref{fig1}(e). Except for the nodal region in $\mathbf{k}$ space, Andreev reflection can be completely suppressed, because the d-wave altermagnet with an even-parity symmetry yields the same spin for both $\mathbf{k}$ and $\mathbf{-k}$ while Andreev reflection requires electrons from $|\mathbf{k}, \uparrow>$ and $|\mathbf{-k}, \downarrow>$ to form a Cooper pair across the interface. Consequently, Andreev reflection can probe a new spin polarization value defined as $P_{\mathrm{AR}} =\frac{\int d^3k\, |N_{\mathbf{k}, \uparrow} - N_{-\mathbf{k}, \downarrow}|}{\int d^3k\, (N_{\mathbf{k}, \uparrow} + N_{-\mathbf{k}, \downarrow})}$. In even-parity altermagnets (e.g., with d or g-wave–type spin textures), the spin-resolved densities of states satisfy \(N_{\mathbf{k},\sigma}=N_{-\mathbf{k},\sigma}\), immediately yielding $P_{\mathrm{AR}} =P_{\mathbf{k}}=\frac{\int d^3k\, |N_{\mathbf{k}, \uparrow} - N_{-\mathbf{k}, \downarrow}|}{\int d^3k\, (N_{\mathbf{k}, \uparrow} + N_{-\mathbf{k}, \downarrow})}$.
We would thus argue that such a $\mathbf{k}$- and spin-dependent Andreev reflection enables it to probe the new type of $\mathbf{k}$-dependent spin polarization $P_k$ in altermagnets. Despite of this theoretical consideration, Andreev reflection in altermagnets has not yet been experimentally explored to confirm its validity.

In this letter, we have applied mechanical point-contact spectroscopy on CrSb single crystals in three major crystallographic orientations to investigate its characteristic behaviors of momentum-dependent spin polarization $P_k$. Our PCS results have unambiguously observed a finite $P_k$ among different planes with average values ranging from 73.4$\%$ for the (0 0 0 1) plane, and 67.9$\%$ for ($\overline{1}$ $\overline{1}$  2 0), to 61.9$\%$ for (1 0 $\overline{1}$ 0), respectively. In addition, the obtained $P_k$ values keep constant below the Nb superconducting temperature, supporting its intrinsic nature. Our work has confirmed the existence of momentum-dependent spin polarization and domain structures in the altermagnetic CrSb, offering a new paradigm to probe the intrinsic nature of altermagnets.

High-quality CrSb single crystals were synthesized by the tin-flux method as described elsewhere \cite{CrSbgrow}. Three major crystallographic planes (0 0 0 1), ($\overline{1}$ $\overline{1}$  2 0), and (1 0 $\overline{1}$ 0) were prepared and their orientations were confirmed by the Laue methods (Refer to the supplementary material for more detailed information \cite{SUPPLEMENTARYMATERIALS}\nocite{chenunifiedformalismandreev2012, strijkersAndreevReflectionsMetal2001c}). As shown in Fig. \ref{fig2}(b), the hexagonal prism represents the crystal structure of CrSb and three major crystallographic planes are highlighted in different colors: the blue plane corresponds to (0 0 0 1), the orange is ($\overline{1}$ $\overline{1}$  2 0), and the green is (1 0 $\overline{1}$ 0), respectively. Samples were carefully polished to ensure a well-defined smooth surface for PCS, and superconducting niobium (Nb) tips were prepared by electrochemical etching with a 12 mol/L sodium hydroxide (NaOH) solution. Point-contacts were established by precisely engaging a sharp Nb tip on the CrSb sample surface with Attocube nano-positioners at low temperatures. The contact conductance G was measured by the conventional lock-in technique in a quasi-four-probe configuration and its values as a function of bias voltage, G(V), were recorded. The CrSb sample and Nb tip were cooled down to 0.3 K and 1.5 K by the Oxford $^3$He and $^4$He refrigerator, respectively, for low-temperature measurements.


\begin{figure}
  \centering
  \includegraphics[angle=0,width=0.48 \textwidth]{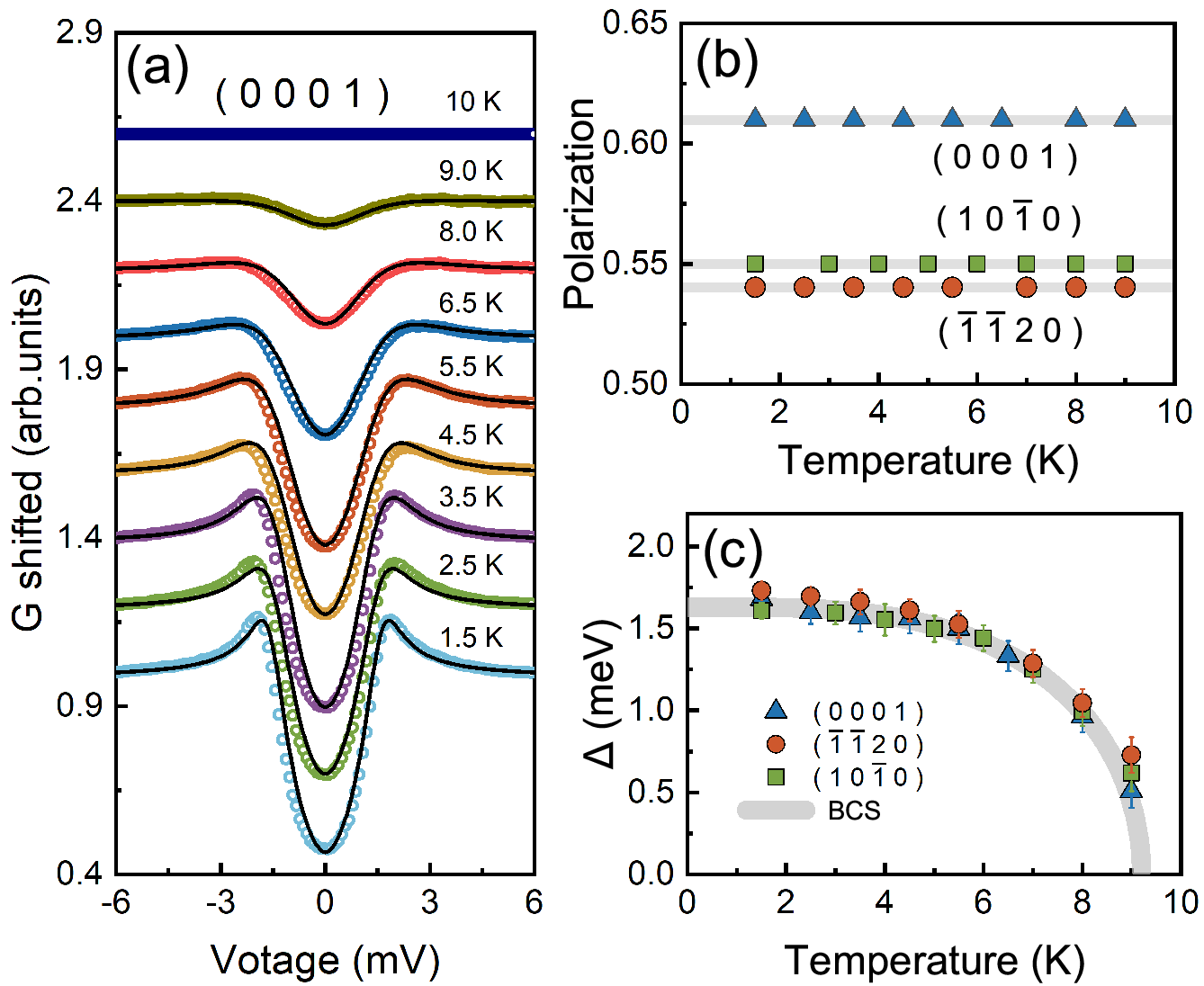}
  \caption{
  \textbf{(a)} Conductance G(V) curves as a function of temperature for a point-contact on the (0 0 0 1) plane.
  \textbf{(b)} The extracted spin-polarization keeps constant with temperature for point-contacts in three major crystallographic planes.
  \textbf{(c)} Temperature dependence of the extracted Nb superconducting gap $\Delta$ for point-contacts in three directions. The gray thick line is the expected BCS temperature behavior, and the extracted gaps $\Delta$ from all three major planes show a good agreement with the expectation.
  }
  \label{fig3}
	\vspace{-10pt}
\end{figure}

We have systematically measured the Andreev reflection spectra on CrSb single crystals along three major crystallographic planes, and the representative conductance spectra $G(V)$ obtained on the (0 0 0 1) plane at 1.5\,K are shown in Fig.~\ref{fig2}(c), while Figs. ~\ref{fig2}(d) and (e) display the corresponding spectra measured on the ($\overline{1}$ $\overline{1}$ 2 0) and (1\,0\,$\overline{1}$\,0) planes at 0.3\,K, respectively. All the conductance curves exhibit a characteristic double-peak structure with a pronounced dip at zero bias, indicative of a finite spin polarization in CrSb. To quantitatively determine its value, a modified BTK model was employed to fit the data with four parameters \cite{chenunifiedformalismandreev2012}: the momentum-dependent spin polarization $P_k$, barrier strength $Z$, smearing factor $\Gamma$, and the superconducting gap $\Delta$. The fitting curves (solid lines) show excellent agreement with the experimental conductance curves (open circles), confirming the reliability of the extracted parameters. Both the inelastic scattering rate $\Gamma$ and the barrier strength $Z$ fall within a relatively narrow range among different contacts (Refer to the supplementary material for more detailed information \cite{SUPPLEMENTARYMATERIALS}), demonstrating the reproducibility and consistency of our MPCS measurements. Furthermore, the dependence of $P_k$ on $Z$ for all three crystallographic orientations is plotted in Fig.~\ref{fig2}(f)-(h). As typically observed in point-contact spectroscopy on ferromagnets~\cite{strijkersAndreevReflectionsMetal2001c}, an increase in $Z$ leads to a reduction in $P_k$ due to spin-mixing effects at the interface. By linearly extrapolating to the $Z \to 0$ limit, we obtain the intrinsic $P_k$ values of approximately 73.4\% for the (0\,0\,0\,1) plane, 67.9\% for the ($\overline{1}$\,$\overline{1}$\,2\,0) plane, and 61.9\% for the (1\,0\,$\overline{1}$\,0) plane.

In addition, the temperature evolution of PCS spectra and spin-polarizations were double-checked, and Fig. \ref{fig3}(a) shows the G(V) curves for the (0 0 0 1) plane with increased temperatures from 1.5 to 10 K, where the zero-bias dip and superconducting peaks become smeared and finally get diminished. At 10 K, the Nb tip becomes in the normal state and the G(V) curve becomes flat. Similar behaviors have been observed for point-contacts on other two orientations (Refer to the supplementary material for more detailed information \cite{SUPPLEMENTARYMATERIALS}). Fig. \ref{fig3}(b) plots the extracted $P_k$ values at different temperatures for all three crystallographic planes, and the spin-polarization remains the same below the Nb superconducting temperature, strongly supporting the viability of our PCS method. Furthermore, the extracted Nb superconducting gap values as a function of temperature are also plotted in Fig. \ref{fig3}(c) and they exactly follow the BCS temperature behavior as shown by the grey area \cite{bardeentheorysuperconductivity1957a}. We note that the Nb gaps obtained here are very close to the reported value of 1.6 meV in literature, while previous studies on ferromagnetic materials often observed a suppressed Nb gap probably owing to the negative proximity effect on the superconducting Nb \cite{howladerstrongspindepolarization2020}. The absence of such a gap suppression in PCS implies that the exotic altermagnetism in CrSb should maintain a zero net magnetization with an alternating spin polarization in the momentum space so to preserve the intrinsic superconducting gap in Nb, which promises an unparalleled advantage of CrSb for future device applications.

\begin{figure}[h]
  \centering
  \includegraphics[angle=0,width=0.45\textwidth]{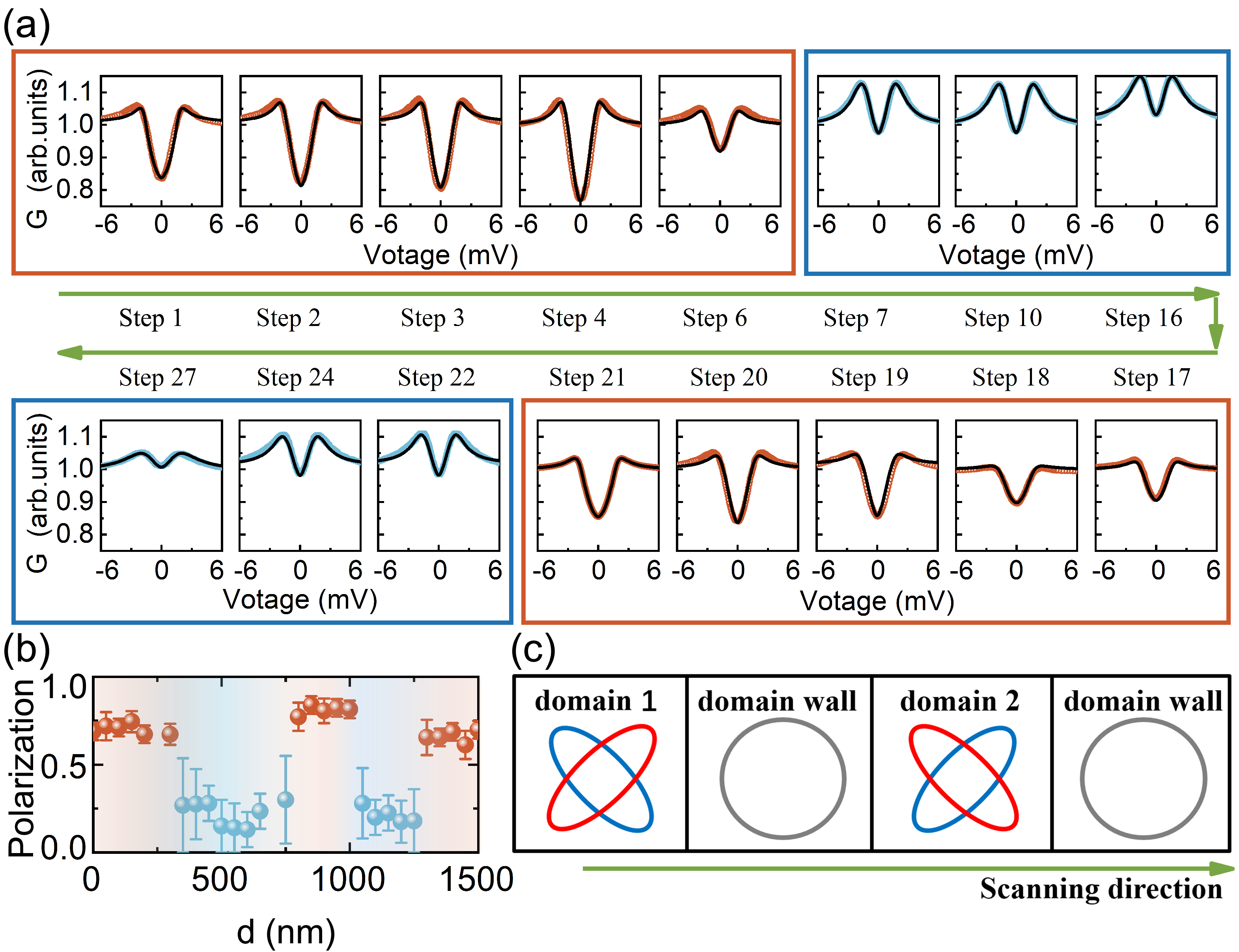}
  \vspace{0pt}
  \caption{\ 
  \textbf{(a)} Spatial evolution of MPCS spectra measured at consecutive positions. 
  \textbf{(b)} The extracted spin polarization values as a function of displacement suggest the existence of domains and domain walls. 
  \textbf{(c)} Schematic illustration of the spin texture inside neighboring domains and the domain wall in between.}
  \label{fig4}
  \vspace{-5pt}
\end{figure}

To explore the probable altermagnetic domains, we have performed spatial line-scans of the conductance spectra with nanopositioners to achieve a controllable displacement on the sample surface at low-temperatures. The Nb tip was repetitively engaged on the CrSb sample and G(V) conductance curves were acquired at each consecutive position. As shown in Figure \ref{fig4} (a), two distinct sets of G(V) curves can be observed during the scan, where some regions show a consistent conductance dip and smeared double-peak feature, signaling the altermagnetic domain with large $P_k$ values, while the spectra in other regions have a shallow zero-bias dip and pronounced double-peak, favoring the existence of domain walls with negligible spin polarizations as illustrated in Fig. \ref{fig4} (c). The extracted $P_k$ values as a function of relative displacements are plotted in Fig. \ref{fig4} (b), where the orange- and blue-shaded areas denote domains and domain-walls, respectively. Importantly, the obtained polarization values inside neighboring domains are consistently high, demonstrating the advantage of MPCS to probe the local spin imbalance instead of the spatially-averaged spin as in ARPES. The reduced $P_k$ values in domain walls suggest a probable breakdown of the long-range altermagnetic order and the recovery of spin degeneracy at $\mathbf{k}$ within the area. Based on the calibrated displacement for each step of the nanopositioner at low temperatures, we estimate that the typical altermagnetic domain size is roughly 250-500~nm while the domain wall is about 250~nm thick, comparable to that reported in MnTe \cite{aminnanoscaleimagingcontrol2024}. MPCS can thus serve as a powerful tool for probing both the k-dependent spin polarization $P_k$ and spatial domains in altermagnetic materials.

\begin{figure}
  \includegraphics[angle=0,width=0.45\textwidth]{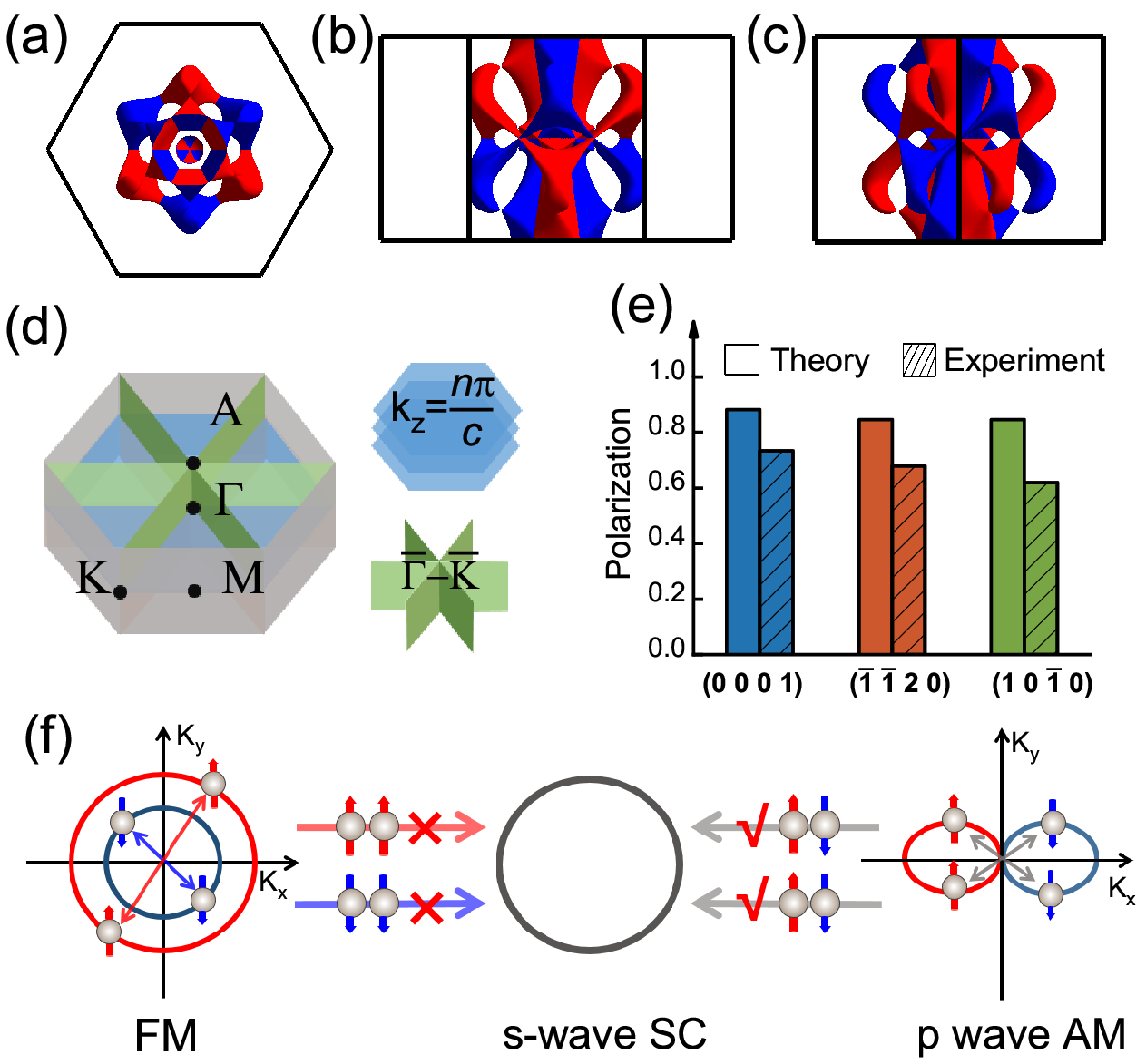}
  \caption{\
  \textbf{(a)} Top view of the calculated Fermi surface of CrSb.
  \textbf{(b)} Side view of the Fermi surface along the $\Gamma$--M direction.
  \textbf{(c)} Side view of the Fermi surface along the $\Gamma$--K direction. The topology and spin texture are characteristic of the g-wave altermagnetic symmetry.
  \textbf{(d)} Three-dimensional Brillouin zone (BZ) shows the high-symmetry points and the nodal planes are located at $k_z = n\pi/c$ ($n$ is an integer) and along the $\overline{\Gamma}$--$\overline{K}$ directions. 
	\textbf{(e)} Comparison between calculated and MPCS-determined momentum-dependent spin polarizations along different crystallographic directions. 
  \textbf{(f)} Schematic illustration of Andreev reflection at the FM/SC and odd-parity AM/SC interface. Red and blue colors denote spin-up and spin-down states, respectively. Electrons at $\mathbf{k}$ and $-\mathbf{k}$ in FM share the same spin, resulting in a total suppression of Andreev reflection at the FM/SC interface, whereas the opposite spins in a p-wave altermagnet enable Andreev reflection at the AM/SC interface. 
  }
    \label{fig5}
		\vspace{-10pt}
\end{figure}

In order to better understand the results above, we have also performed first-principles based simulations. By fitting the first-principles electronic structure to a tight-binding Hamiltonian using Cr-3d and Sb-5p orbitals with symmetrized maximally projected Wannier functions \cite{method:mlwf,ZHI2022108196}, we calculate the spin polarized Fermi surfaces with a g-wave symmetry as in Fig.\ref{fig5}(a)-(c) and spin-momentum resolved density of states close to the Fermi level. Employing full ballistic limit approximation \cite{IMazin99PRL}, we approximated the integration of $\mathbf{k}$-dependent spin polarization along the probing direction $\hat{\mathbf{n}}$ by 
\begin{equation*}
  P(\hat{\mathbf{n}})= \frac{\sum_{i,\sigma} \int d^3\mathbf{k} v_{i\sigma\mathbf{k}} |\delta(\mu-\epsilon_{i\sigma\mathbf{k}})-\delta(\mu-\epsilon_{i\bar{\sigma},-\mathbf{k}})| } {\sum_{i,\sigma} \int d^3\mathbf{k}  v_{i\sigma\mathbf{k}} (\delta(\mu-\epsilon_{i\sigma\mathbf{k}})+\delta(\mu-\epsilon_{i\bar{\sigma}-\mathbf{k}})) }
\end{equation*}
where the anisotropy due to the band specific transmission coefficient and interface barrier are ignored, and $v_{i\sigma\mathbf{k}}=\hat{\mathbf{n}}\cdot\mathbf{v}_{i\sigma\mathbf{k}}$ selects the relevant $i{\mathbf{k}}$ states, $\sigma$ indicates the spin index, which enumerates $\uparrow$ and $\downarrow$, and $\bar{\sigma}$ indicates the opposite spin of $\sigma$. The results remain essentially the same even if the diffusive limit formula is employed, suggesting that the crucial ingredient here is the momentum dependence of the spin polarization. We note that the $\mathbf{k}$-space selection rule imposed by the Andreev reflection process is local in the reciprocal space, thus the integration cannot be recast into integration of energy. Our calculations give an averaged spin-polarization of 87.4$\%$ for the (0 0 0 1) plane, 85.6$\%$ for ($\overline{1}$ $\overline{1}$  2 0) and 85.4$\%$ for (1 0 $\overline{1}$ 0) plane as shown in Fig. \ref{fig5}(e). In addition, since our estimation of spin polarization completely ignores transmission coefficient and other details, the calculated $P(\hat{\mathbf{n}})$ is always larger than the observed $P_k$ values. A more realistic model is highly desirable to explore the discrepancy.


We stress that a special care should be given to distinguish the global, momentum-dependent and Andreev spin polarizations ($P$, $P_k$ and $P_{AR}$). If we consider a ferromagnet case with a partial global polarization $0 < P < 1$ as shown in the left panel of Fig.~\ref{fig5}(f), the even-parity nature of the spin splitting ($P_k=1$ with $N_{\mathbf{k},\sigma}=N_{-\mathbf{k},\sigma}$) totally suppresses the Andreev reflection due to the same spins for $\mathbf{k}$ and $\mathbf{-k}$ and results in a full Andreev polarization $P_{AR} = P_k = 1$. Meanwhile, if the unconventional magnet owns an odd-parity symmetry such as $p$-wave in the right panel of Fig.~\ref{fig5}(f), the global polarization vanishes with $P=0$, while the momentum-dependent polarization $P_{k}\approx 1$ ignoring the nodal region. However, the opposite spins for $\mathbf{k}$ and $\mathbf{-k}$ would lead to a null Andreev polarization $P_{AR} = 0$, where Andreev reflection can still be observed.

In summary, we have proposed a new paradigm to define the momentum-dependent spin-polarization $P_k$ with Andreev reflection process and employed point-contact spectroscopy with superconducting Nb tips to probe $P_k$ in the altermagnetic candidate CrSb along three major crystallographic planes, yielding an averaged value of 73.4$\%$ for the (0 0 0 1) plane, 67.9$\%$ for ($\overline{1}$ $\overline{1}$  2 0), and 61.9$\%$ for (1 0 $\overline{1}$ 0), respectively. A spatial line-scan of the G(V) curves on the sample surface reveal the existence of altermagnetic domains with a characteristic domain size of 250-500~nm and domain-wall width about 250 nm. Our results not only confirm Andreev reflection as a powerful probe for momentum-dependent spin polarization $P_k$ in altermagnets, but also support the altermagnetic CrSb can serve as an ideal platform for potential applications.

\begin{acknowledgments}
This work was supported by the National Key R\&D Program of China (Grant No. 2025YFA1411500 and 2022YFA1402200), the National Natural Science Foundation of China (Grant No. 12574147, 12274364, and 12174333). Y.F.G. acknowledges the open research fund of Beijing National Laboratory for Condensed Matter Physics (2023BNLCMPKF002). Y.X. was further supported by the National Natural Science Foundation of China (Grant Nos. 12550401 and 12374163) and Zhejiang Provincial Natural Science Foundation of China (Grant No. LR26A040003).
\end{acknowledgments}

\paragraph{Data availability} The data that support the findings of this article are not publicly available. The data are available from the authors upon reasonable request.


%

\end{document}